# Authentication over Noisy Channels

Lifeng Lai[*], Hesham El Gamal[†], H. Vincent Poor[‡]
[*][‡] Department of Electrical Engineering
Princeton University, Princeton, NJ 08512, USA
Email: {llai,poor}@princeton.edu
[†] Department of Electrical and Computer Engineering
Ohio State University
Columbus, OH 43202, USA
Email: helgamal@ece.osu.edu

*Abstract*— In this work, message authentication over noisy channels is studied. The model developed in this paper is the authentication theory counterpart of Wyner's wiretap channel model. Two types of opponent attacks, namely impersonation attacks and substitution attacks, are investigated for both single message and multiple message authentication scenarios. For each scenario, information theoretic lower and upper bounds on the opponent's success probability are derived. Remarkably, in both scenarios, lower and upper bounds are shown to match, and hence the fundamental limit of message authentication over noisy channels is fully characterized. The opponent's success probability is further shown to be smaller than that derived in the classic authentication model in which the channel is assumed to be noiseless. These results rely on a proposed novel authentication scheme in which key information is used to provide simultaneous protection again both types of attacks.

## I. INTRODUCTION

There are two fundamental primitives for any security systems: 1) *secure transmission*, to ensure that the message is received only by the legitimate *receiver*; 2) *authentication*, to ensure that the received message truly comes from the acclaimed *transmitter*.

Secure transmission has been investigated under two different models. In the model developed by Shannon [1], transmissions are assumed to be noiseless; and the source and intended destination use a common secret key $K$ to encrypt and decrypt the message $M$. Transmission is said to be *perfectly* secure, if the signal received at the opponent does not provide it with any information about $M$. Shannon proved that one needs $H(K) \geq H(M)$ to achieve perfect security. Taking transmission noise into consideration, Wyner developed the wiretap channel [2], in which the transmitter exploits the two different noise processes at the receiver and opponent to transmit information securely. Csiszár and Körner [3] generalized this model and characterized the capacity of the Discrete Memoryless Channel (DMC) with security constraints.

Authentication theory with a noiseless transmission model, which is shown in Figure 1, was developed by Simmons [4]. In this model, the source $S$ and the receiver $R$ share a secret key $K$, which is used to identify the transmitter. When the transmitter intends to send message $M$, it transmits $W =$

This research was supported by the National Science Foundation under Grants ANI-03-38807 and CNS-06-25637.

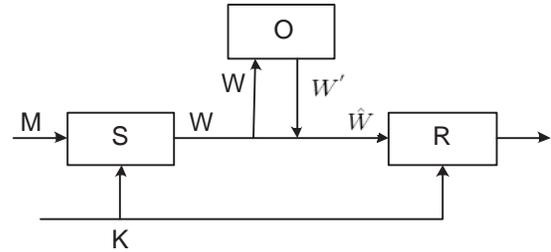

Fig. 1. The authentication channel.

$f(K, M)$ over a noiseless public channel, where $f$ is the encoding function at the source. On receiving $\hat{W}$, which might be different from $W$ due to various attacks from the opponent $O$, the receiver needs to judge whether the message comes from the legitimate transmitter or not. If the receiver accepts the message (i.e., the receiver believes that the signal is authentic), the receiver then gets an estimate of the source message $M$; otherwise, it rejects the message. The opponent gets a perfect copy of $W$ and can perform the following two types of attacks. The first one is called an *impersonation attack*, in which the opponent sends $W'$ to the destination before the source sends anything. This attack is successful if $W'$ is accepted by the receiver as authentic. We denote the success probability of this attack by $P_I$. The second attack is called a *substitution attack*, in which after receiving $W$, the opponent modifies it to $W'$ and sends it to the destination. The attack is successful if the receiver accepts $W'$ and decodes this into another source state. We denote the success probability of this attack by $P_S$. Obviously, the opponent will choose the attack that has higher success probability. Hence the success probability $P_D$ of the opponent (i.e., the *cheating probability*) is $P_D = \max\{P_I, P_S\}$.

Lower bounds on $P_I$ and $P_S$ have been developed in [4] and recovered by Maurer [5] from a hypothesis testing point of view. In particular, it was shown that $P_I \geq 2^{-I(K;W)}$ and $P_S \geq 2^{-H(K|W)}$. One can easily identify a tradeoff between $P_I$ and $P_S$. To minimize the probability of a successful impersonation attack, the transmitted ciphertext, from the legitimate source, must contain a sufficient amount of information about the secret key in order to convince the legitimate receiver that the transmitted message comes from the source. That is $I(K;W)$ should be large, which unfortunately decreases $H(K|W)$. Hence, the attacker can take advantage of the leaked information over its noiseless channel (contained in $W$) to increase the probability of a successful substitution attack. In fact, the strategy that minimizes the lower bound on $P_D = \max\{P_I, P_S\}$ is to use half of the key information to protect against the impersonation attack and the other half of the key information to protect against the substitution attack, which gives $P_D \geq 2^{-H(K)/2}$. These bounds are of a negative nature, since they only give lower bounds for the cheating probability. There is no upper-bound available in the literature, partly due to the fact that usual bounding techniques such as Jensen's inequality and the log-sum inequality are not applicable here. We will elaborate on this point in the sequel.

Simmons's model was developed under a noiseless transmission model. However, since physical transmission systems are noisy, common practice is to use channel coding to convert the noisy channel into a noiseless one, and then to design an authentication code on top of the channel coding. Liu and Boncelet [6], [7] also considered the situation in which the channel coding is not perfect, and hence there are some residual errors induced by the channel. The conclusion of these papers is that channel noise is detrimental to authentication, since it will cause the receiver to reject authentic messages from the transmitter.

In this paper, we take an alternative view of the transmission noise and design the channel coding and authentication scheme jointly. We show that by doing so, one can exploit the noise to lower the cheating probability of the opponent. More specifically, we derive both a lower bound and an upper-bound on the cheating probabilities of authentication schemes over noisy channels. We show that these two bounds coincide, and are smaller than the lower-bound on the cheating probability when the channel is assumed to be noiseless. In particular, we show that $P_D = 2^{-H(K)}$, thus all the key information can be used to protect against the substitution attack and the impersonation attack simultaneously. We also study the authentication of multiple messages using the same key $K$, and show that all the key information can be used to protect against all the attacks simultaneously.

The rest of the paper is organized as follows. In Section II, we introduce the model. In Section III, we discuss the single message authentication scenario. We then analyze the authentication of multiple message using a same key in Section IV. Finally, in Section V, we offer some conclusions.

## II. MODEL

Throughout this paper, upper-case letters (e.g., $X$) denote random variables, lower-case letters (e.g., $x$) denote realizations of the corresponding random variables, and calligraphic letters (e.g, $\mathcal{X}$) denote finite alphabet sets over which corresponding variables range. Also, upper-case boldface letters (e.g., $\mathbf{X}$) denote random vectors and lower-case boldface letters (e.g., $\mathbf{x}$) denote realizations of the corresponding random vectors.

Figure 2 shows the model under consideration. The model differs from Simmons's model only in that the transmission channel is noisy. More specifically, we consider the DMC and assume that when the transmitter sends $\mathbf{x}$, the receiver receives $\mathbf{y}$ with probability

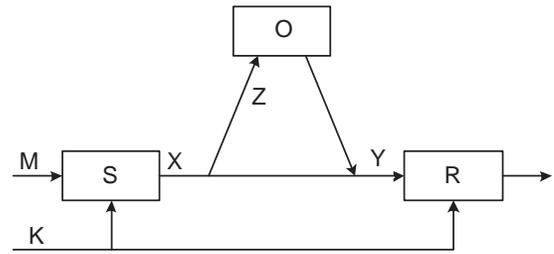

Fig. 2. The authentication channel.

$$P_{Y|X}^n(\mathbf{y}|\mathbf{x}) = \prod_{j=1}^{n} P(y|x)$$

and the opponent receives $\mathbf{z}$ with probability

$$P_{Z|X}^n(\mathbf{z}|\mathbf{x}) = \prod_{j=1}^{n} P(z|x).$$

Here $P(y|x)$ and $P(z|x)$ denote the channel transition probabilities, while $x, y$ and $z$ range through finite sets $\mathcal{X}$, $\mathcal{Y}$ and $\mathcal{Z}$, respectively. In order to derive more general bounds, we assume that the channel between the opponent and receiver is noiseless, and that the opponent can send anything over this channel. Note that this assumption does not incur any loss of generality, and actually gives the opponent advantages, since any noisy channel can be simulated by this noiseless channel by simply randomizing the transmitted signal.

To identify the transmitter, we assume that the source and the destination have a common secret key $K$ ranging from a set $\mathcal{K}$ having $|\mathcal{K}|$ possible values. To transmit the message $M$, the source uses a stochastic encoding function $f$ to convert the message and key into a length $n$ vector $\mathbf{X}$, i.e., $\mathbf{X} = f(K, M)$. Upon receiving $\mathbf{Y}$, which may come from either the source or the opponent, the destination uses a decoding function $g$ to judge whether the message is authentic or not. If the signal is deemed authentic, then the destination recovers the message $\hat{M} = g(\mathbf{Y}, K)$; otherwise the destination sets $\phi = g(\mathbf{Y}, K)$. We require the condition that, if the signal is authentic, the decoding error probability at the destination must approach zero as the length of the code increases, i.e., for any $\epsilon > 0$, there is a positive integer $n_0$, such that when $n \geq n_0$, we have

$$P_e = \Pr\{g(\mathbf{Y}, K) \neq M | \mathbf{Y} \text{ comes from } \mathbf{X}\} \leq \epsilon.$$

The error probability $P_e$ consists of two parts: $P_1$ and $P_2$, where $P_1$ is the probability of a miss, which is the probability that the receiver wrongly rejects an authentic message, and $P_2$ is the probability that the decoder correctly accepts the signal as being authentic but incorrectly decodes it.

The opponent is assumed to be aware of the system design, except for the particular realizations $k$ and $m$ of the key $K$ and message $M$. We consider both of the two forms of attack described above. That is, we consider the impersonation attack, in which the opponent sends codeword **X** to the receiver before the transmitter sends anything. Such an attack is successful if **X** is accepted as authentic by the receiver, and we denote this probability of success as $P_I$ as noted above. We also consider the substitution attack, in which the opponent blocks the transmission of the main channel while receiving **Z**. After that, the opponent modifies the signal and transmits it to the receiver. This attack is considered to be successful if the modified signal is accepted as authentic by the receiver and is decoded into $m'$ that is not equal to the original message $m$. Again, the success probability of this attack is denoted by $P_S$.

### III. Authentication of A Single Message

#### A. The Wiretap Channel

We begin by reviewing some results related to the wiretap channel introduced in [2]. The wiretap channel is defined by two DMCs $\mathcal{X} \to (\mathcal{Y}, \mathcal{Z})$, where $\mathcal{X}$ is the input alphabet from the transmitter, $\mathcal{Y}$ is the output alphabet at the legitimate receiver and $\mathcal{Z}$ is the output alphabet at the wiretapper. In the wiretap channel, the wiretapper is assumed to be passive, and the goal is to transmit information to the destination while preventing information leakage to the wiretapper. More specifically, to send a message $M \in \mathcal{M}$, the transmitter sends $\mathbf{X} = f(M)$, where $f$ is a stochastic encoder. After receiving **Y**, the destination obtains an estimate $\hat{M} = g(\mathbf{Y})$. A perfectly secure rate $R_s$ is said to be achievable if there exist $f$ and $g$, such that for reach $\epsilon > 0$, there is a positive integer $n_0$, such that $\forall n > n_0$

$$|\mathcal{M}| \geq 2^{nR_s} \quad (1)$$
$$\Pr\{\hat{M} \neq M\} \leq \epsilon, \text{ and} \quad (2)$$
$$\frac{1}{n} I(M; \mathbf{Z}) \leq \epsilon. \quad (3)$$

The perfect secrecy capacity $C_s$ is defined to be the supremum of the set of $R_s$ values that satisfy the conditions (1) - (3). It is proved in [3] that the perfect secrecy capacity is given by

$$C_s = \max_{U \to X \to YZ} [I(U;Y) - I(U;Z)],$$

where $U$ is an auxiliary random variable satisfying the Markov chain relationship $U \to X \to YZ$.

The source-wiretapper channel is said to be less noisy than the main channel, if for all possible $U$ that satisfy the above Markov chain relationship, one has $I(U;Z) > I(U;Y)$. We can see that the perfect secrecy capacity is nonzero unless the wiretapper channel is less noisy than the main channel.

#### B. Authentication Scheme

We use the wiretap channel to perform authentication. More specifically, if the wiretapper channel is not less noisy than the main channel, there exists an input distribution $P_X$ such that $I(X;Y) - I(X;Z) > 0$. For a given key size $|\mathcal{K}|$, there exists a positive integer $n_0$, such that $\forall n \geq n_0$,

$$\exp\{n(I(X;Y) - I(X;Z))\} > |\mathcal{K}|.$$

In our transmission scheme, we separate the transmission of information and key. The source first sends the message $M$ using a code for the wiretap channel, and then sends the key $K$ using the same code book. After receiving these signals, the destination obtains an estimate $\hat{M}$ of the message and a separate estimate $\hat{K}$ of the key. If $\hat{K} = K$, the receiver accepts the message to be authentic; otherwise it rejects the message.

For an impersonation attack, the optimal strategy for the opponent is to choose the key that has the largest probability of being accepted by the receiver, i.e.,

$$P_I = \max_{k' \in \mathcal{K}} \left\{ \sum_{k \in \mathcal{K}} P(k) \gamma(k, k') \right\},$$

where $\gamma(k, k')$ is an indicator function that equals 1 if $k'$ is accepted as authentic, and equals 0 in other cases. In our scheme, $\gamma(k, k') = 1$ if $k' = k$; otherwise $\gamma(k, k') = 0$.

For a substitution attack, the optimal strategy for the opponent is to choose $m'$ and $k'$ such that the probability of the message being accepted by the receiver and being decoded into $m' \neq m$, is maximized, i.e.,

$$P_S = \sum_{\mathbf{z}_1, \mathbf{z}_2} P(\mathbf{z}_1, \mathbf{z}_2)$$
$$\max_{m' \in \mathcal{M}, k' \in \mathcal{K}} \left\{ \sum_{m,k} P(m, k | \mathbf{z}_1, \mathbf{z}_2) \gamma(m, k, m', k') \right\}$$
$$= \sum_{\mathbf{z}_1, \mathbf{z}_2} P(\mathbf{z}_1) P(\mathbf{z}_2)$$
$$\max_{m' \in \mathcal{M}, k' \in \mathcal{K}} \left\{ \sum_{m,k} P(m|\mathbf{z}_1) P(k|\mathbf{z}_2) \gamma(m, m') \gamma(k, k') \right\},$$

where $\mathbf{z}_1$ is the signal received for the message part and $\mathbf{z}_2$ is the signal received for the key part. Here $\gamma(m, k, m', k') = 1$ if $m' \neq m$ and $k' = k$, and equals 0 otherwise. The second equality in the above expression is due to the fact that $M$ and $K$ are independent, and thus that $\mathbf{Z}_1$ and $\mathbf{Z}_2$ are also independent.

To simplify the analysis, we first upper-bound $P_S$ as follows

$$P_S = \sum_{\mathbf{z}_1, \mathbf{z}_2} P(\mathbf{z}_1) P(\mathbf{z}_2)$$
$$\max_{m' \in \mathcal{M}, k' \in \mathcal{K}} \left\{ \sum_{m,k} P(m|\mathbf{z}_1) P(k|\mathbf{z}_2) \gamma(m, m') \gamma(k, k') \right\}$$
$$\stackrel{(a)}{\leq} \sum_{\mathbf{z}_1, \mathbf{z}_2} P(\mathbf{z}_1) P(\mathbf{z}_2) \max_{m \in \mathcal{M}, k \in \mathcal{K}} \{P(m|\mathbf{z}_1) P(k|\mathbf{z}_2)\}$$
$$\stackrel{(b)}{\leq} \sum_{\mathbf{z}_1} P(\mathbf{z}_1) \left( \sum_{\mathbf{z}_2} P(\mathbf{z}_2) \max_{k \in \mathcal{K}} \{P(k|\mathbf{z}_2)\} \right)$$
$$= \sum_{\mathbf{z}_2} P(\mathbf{z}_2) \max_{k \in \mathcal{K}} \{P(k|\mathbf{z}_2)\}. \quad (4)$$

In this expression, inequality (a) follows by assuming that $\gamma(m, m') = 1$ and $\gamma(k, , k') = 1$ for $m' = \arg \max_{m \in \mathcal{M}} P(m|\mathbf{z}_1)$ and $k' = \arg \max_{k \in \mathcal{K}} P(k|\mathbf{z}_2)$. If this is not the case, the summation will only be smaller, since $\gamma(x, y)$ is the indicator

function. Inequality (b) follows from the fact that $P(m|\mathbf{z}_1) \leq 1$.

In the sequel, we will use this upper-bound, and hence we can ignore the message transmission part $\mathbf{z}_1$. Consequently, we write $\mathbf{z}_2$ as $\mathbf{z}$ for the sake of simplicity of notation.

After receiving $\mathbf{Z}$, the opponent gains an amount $I(K;\mathbf{Z})$ of information about the key, and thus can use this information to choose $k$ that maximizes $P(k|\mathbf{z}_2)$. From (3), we have that

$$I(K;\mathbf{Z}) \leq n\epsilon. \tag{5}$$

The inequality in (5) is not enough to analyze (4) for the following two reasons. First, though $\epsilon$ is small, $n\epsilon$ can go to infinity as $n$ grows, and hence the opponent may eventually gain a sufficient amount of information about the key. This point has been pointed out in [8]–[10]. The second reason is that there is a maximization in the summand in (4), which means that we need to consider the worst case scenario, whereas $I(K;\mathbf{Z})$ is an average quantity. Actually, this fact is exploited in [4], [5] to derive the lower bounds by replacing this maximization with an averaging, which readily gives us a lower bound and is more amenable to analysis.

In this paper, we borrow techniques from [10], [11] to analyze this term.

*C. Bounds*

We begin with some definitions. Let $\mathcal{C}$ be a codebook for the wiretap channel, and let $\tilde{P}(\mathbf{x},\mathbf{z})$ be the joint distribution on $\mathcal{C} \times \mathcal{Z}^n$. We denote by $Q(\mathbf{z})$ the marginal distribution of $\mathbf{z}$ when the input distribution is limited to $\mathcal{C}$, and by $P(\mathbf{x}|\mathbf{z}) = \tilde{P}(\mathbf{x},\mathbf{z})/Q(\mathbf{z})$ the conditional distribution of $\mathbf{x}$ given $\mathbf{z}$.

Let $\{\mathcal{C}_1,\cdots,\mathcal{C}_N\}$ be a partition of $\mathcal{C}$, and denote this partition as a mapping, i.e., $f : \mathcal{C} \to \{\mathcal{C}_1,\cdots,\mathcal{C}_N\}$. Also denote by $Q_j$ the conditional distribution of $\mathbf{z}$ when the input distribution is uniform on $\mathcal{C}_j$, i.e.,

$$Q_j(\mathbf{z}) = \sum_{\mathbf{x} \in \mathcal{C}_j} \tilde{P}(\mathbf{x},\mathbf{z})/P(\mathcal{C}_j).$$

Define $d_{av}(f) = \sum_{j=1}^{N} P(\mathcal{C}_j) d(Q_j, Q)$, with

$$d(Q_j, Q) = \sum_{\mathbf{z} \in \mathcal{Z}^n} \Big| Q_j(\mathbf{z}) - Q(\mathbf{z}) \Big|.$$

Here $d(Q_j, Q)$ is the $\mathcal{L}_1$ distance between the two distributions $Q_j$ and $Q$. When $d(Q_j, Q)$ is zero, the opponent cannot distinguish between the uniform input distributions on $\mathcal{C}_j$ and $\mathcal{C}$ by observing only the channel output.

Intuitively, if there exists a set $\mathcal{C}$ and a corresponding partition $f$ such that $d_{av}(f)$ is arbitrarily small, the receiver gains no information about the subset $\mathcal{C}_j$ from which the transmitted codeword $\mathbf{x}$ comes, given the channel output $\mathbf{z}$.

We can rewrite $d_{av}(f)$ as follows

$$\begin{aligned} d_{av}(f) &= \sum_{j=1}^{N} \sum_{\mathbf{z} \in \mathcal{Z}^n} \Big| P(\mathcal{C}_j) Q_j(\mathbf{z}) - P(\mathcal{C}_j) Q(\mathbf{z}) \Big| \\ &= \sum_{\mathbf{z} \in \mathcal{Z}^n} Q(\mathbf{z}) d(\mathbf{z}), \end{aligned}$$

with

$$d(\mathbf{z}) = \sum_{j=1}^{N} \Big| P(\mathcal{C}_j|\mathbf{z}) - P(\mathcal{C}_j) \Big|.$$

Here $d(\mathbf{z})$ is the $\mathcal{L}_1$ distance between uniform distribution and conditional distribution of the key after observing $\mathbf{z}$ at the opponent.

We need the following lemma from [10].

*Lemma 1 ( [10]):* Consider a wiretap channel $\mathcal{X} \to (\mathcal{Y}, \mathcal{Z})$, and choose $\delta > 0$. Suppose $\mathcal{T}_P \subset \mathcal{X}^n$ is a type class with $P(x)$ bounded away from 0, and such that $I(X;Y) > I(X;Z) + 2\delta$. Then, there exist a codebook $\mathcal{C}$ with size $|\mathcal{C}| = \exp\{n(I(X;Y) - \delta)\}$, drawn from $\mathcal{T}_P$, and equal-size disjoint subsets $\mathcal{C}_1, \cdots, \mathcal{C}_N$ of $\mathcal{C}$ with

$$N \leq \exp\{n(I(X;Y) - I(X;Z) - 2\delta)\},$$

such that $\mathcal{C} = \bigcup_{i=1}^{N} \mathcal{C}_i$ is the codeword with exponentially small average probability of error for the main channel $\mathcal{X} \to \mathcal{Y}$. Moreover, the partition function $f : \mathcal{C} \to \{1, \cdots, N\}$ of $\mathcal{C}$ with $f^{-1}(i) = \mathcal{C}_i, i = 1, \cdots, N$ has exponentially small $d_{av}(f)$ for the distribution $\tilde{P}_C$ defined on $\mathcal{C} \times \mathcal{Z}^n$ by

$$\tilde{P}_C(\mathbf{x},\mathbf{z}) = \frac{1}{|\mathcal{C}|} P(\mathbf{z}|\mathbf{x}), \mathbf{x} \in \mathcal{C}, \mathbf{z} \in \mathcal{Z}^n.$$

*Proof:* Please see [10]. ∎

Our main result is the following theorem.

*Theorem 1:* If the source-wiretapper channel is not less noisy than the main channel, then $P_I = P_S = 2^{-H(K)}$, and hence, $P_D = 2^{-H(K)}$.

*Proof:* (Sketch) For the lower-bound, the opponent can guess the value of the key. If the guess is correct, the opponent can invoke any attack and the attack will be successful. The probability that the opponent guesses the value of key correctly is $2^{-H(K)}$. This provides a lower bound. We outline the proof of a tight upper-bound in the following. If the source-wiretapper channel is not less noisy than the main channel, there exists an input distribution such that the secrecy rate is larger than zero. We generate a codebook for the wiretap channel according to this input distribution and transmit the message and key separately using this codebook. To bound the success probability of the substitution attack, we first bound the 'max' sign in (4) with $d(\mathbf{z})$. We then link $d_{av}(f)$ to the mutual information leaked to the opponent. Using the fact that the mutual information leakage in the wiretap channel can be arbitrarily small if the secrecy capacity is nonzero, we obtain an upper-bound for the success probability of the substitution attack that is arbitrarily close to $2^{-H(K)}$. The optimal strategy for the impersonation attack of the opponent is to guess the value of the key, hence the success probability of the impersonation attack is bounded by $2^{-H(K)}$. ∎

## IV. AUTHENTICATION OF MULTIPLE MESSAGES

In this section, we consider the situation in which the same key $K$ is used to authenticate a sequence of $J$ messages. We use the same scheme as for the single message case. That is, we send the message and the key separately for each packet using a code for the wiretap channel. Let $P_{I,i}$ be the success probability of the impersonation attack after the opponent has observed $i-1$ transmissions, i.e., the opponent sends codeword $\mathbf{X}_i$ to cheat the destination after observing $\mathbf{Z}_1, \cdots, \mathbf{Z}_{i-1}$. This attack is successful if $\mathbf{X}_i$ is accepted as authentic by the destination. The optimal attack strategy of the opponent is to choose to send the key $k'$ with the largest success probability; that is

$$\begin{aligned} P_{I,i} &= \sum_{\mathbf{z}_1,\cdots,\mathbf{z}_{i-1}} P(\mathbf{z}_1,\cdots,\mathbf{z}_{i-1}) \\ &\qquad \max_{k'\in\mathcal{K}}\left\{\sum_{k\in\mathcal{K}} P(k|\mathbf{z}_1,\cdots,\mathbf{z}_{i-1})\gamma(k,k')\right\} \\ &\leq \sum_{\mathbf{z}_1,\cdots,\mathbf{z}_{i-1}} P(\mathbf{z}_1,\cdots,\mathbf{z}_{i-1}) \\ &\qquad \max_{k\in\mathcal{K}}\{P(k|\mathbf{z}_1,\cdots,\mathbf{z}_{i-1})\}, \end{aligned} \qquad (6)$$

where $\gamma(k,k')$ is the indicator function defined above.

The opponent can also choose to invoke a substitution attack after receiving the $i$th transmission, i.e., it changes the content of the $i$th package and sends it to the destination. The attack is successful if the modified message is accepted as authentic and the destination decodes it into an incorrect source state. On denoting the success probability of this attack to be $P_{S,i}$, we have

$$\begin{aligned} P_{S,i} &= \sum_{\mathbf{z}_{i,0},\mathbf{z}_1,\cdots,\mathbf{z}_i} P(\mathbf{z}_{i,0},\mathbf{z}_1,\cdots,\mathbf{z}_i) \\ &\qquad \max_{m'\in\mathcal{M}, k'\in\mathcal{K}}\left\{\sum_{m,k} P(m,k|\mathbf{z}_{i,0},\mathbf{z}_1,\cdots,\mathbf{z}_i)\gamma(m,k,m',k')\right\}, \end{aligned}$$

where $\mathbf{z}_{i,0}$ is the message part of the $i$th packet. Following the same steps as those in (4), we can bound $P_{S,i}$ as

$$P_{S,i} \leq \sum_{\mathbf{z}_1,\cdots,\mathbf{z}_i} P(\mathbf{z}_1,\cdots,\mathbf{z}_i) \max_{k\in\mathcal{K}}\{P(k|\mathbf{z}_1,\cdots,\mathbf{z}_i)\}. \qquad (7)$$

Note that (6) and (7) have similar forms. Hence, we can derive tight bounds for only one of these attacks. The result for the other attack follows similarly.

Obviously, the opponent will choose the attack that maximizes its cheating probability $P_D$. Bounds for $P_{I,i}$ and $P_{S,i}$ under the noiseless transmission model were derived in [5], which shows that

$$P_D = \max\{P_{I,1},\cdots,P_{I,J}, P_{S,1},\cdots,P_{S,J}\} \geq 2^{-H(K)/(J+1)}.$$

This implies that after several rounds of authentication, the opponent obtains almost all the information about the key and hence can choose an attack having a high success probability.

On the other hand, in the noisy channel model, we show that one can limit the information leaked to the opponent, and thus the success probability of the opponent will not increase even by observing more packets.

*Theorem 2:* For any finite $J$, $P_{I,i} = P_{S,i} = 2^{-H(K)}, i \in \{1,\cdots,J\}$. Hence, $P_D = 2^{-H(K)}$.

*Proof:* (Sketch) For the lower-bound, the opponent can guess the value of the key. If the guess is correct, the opponent can invoke any attack and the attack will be successful. The probability that the opponent guesses the value of key correctly is $2^{-H(K)}$. This provides a lower bound. For a tight upper-bound, we first upper bound the key information leaked to the opponent. We then follow the similar steps as those of the single message authentication case and obtain an upper-bound of the success probability of the substitution attack that is arbitrarily close to $2^{-H(K)}$. Similarly, we obtain an upper-bound for the impersonation attack that is arbitrarily close to $2^{-H(K)}$. ∎

## V. CONCLUSIONS

In this paper, we have studied the problem of message authentication in the presence of channel noise. We have derived information theoretic lower and upper bounds for the success probability of an opponent's impersonation attack and substitution attack in single and multiple message authentication scenarios. We have further shown that the lower and upper bound match, and thus have completely characterized these probabilities. We have further shown that, compared with the classical authentication model in which channel is assumed to be noiseless, the opponent's success probability is largely reduced. We thus have established the utility of channel noise in message authentication applications.

Exploiting other characteristics of channels, such as channel fading, to facilitate message authentication is an interesting avenue for further research. Also of interest is the development of authentication theory for the scenario in which the source and destination possess correlated, but not identical, sequences, which has obvious practical implications.

## REFERENCES

[1] C. E. Shannon, "Communication theory of secrecy systems," *Bell System Technical Journal*, vol. 28, pp. 656–715, Oct. 1949.
[2] A. D. Wyner, "The wire-tap channel," *Bell System Technical Journal*, vol. 54, no. 8, pp. 1355–1387, 1975.
[3] I. Csiszár and J. Körner, "Broadcast channels with confidential messages," *IEEE Transactions on Information Theory*, vol. 24, pp. 339–348, May 1978.
[4] G. J. Simmons, "Authentication theory/coding theory," in *Proceedings of CRYPTO 84 on Advances in Cryptology*, (New York, NY, USA), pp. 411–431, Springer-Verlag Inc., 1985.
[5] U. M. Maurer, "Authentication theory and hypothesis testing," *IEEE Trans. on Information Theory*, vol. 46, pp. 1350–1356, Jul. 2000.
[6] Y. Liu and C. G. Boncelet, "The CRC-NTMAC for noisy message authentication," *IEEE Transactions on Information Forensics and Security*, vol. 1, pp. 517–523, Dec. 2006.
[7] C. G. Boncelet, "The NTMAC for authentication of noisy messages," *IEEE Transactions on Information Forensics and Security*, vol. 1, pp. 35–42, Mar. 2006.
[8] C. H. Bennett, G. Brassard, C. Crepeau, and U. M. Maurer, "Generalized privacy amplification," *IEEE Transactions on Information Theory*, vol. 41, pp. 1915–1923, Nov. 1995.
[9] U. M. Maurer and S. Wolf, "Information-theoretic key agreement: From weak to strong secrecy for free," *Lecture Notes in Computer Science*, vol. 1807, pp. 356–373, 2000.
[10] I. Csiszár, "Almost independence and secrecy capacity," *Problems of Information Transmission*, vol. 32, pp. 40–47, Jan. 1996.
[11] R. Ahlswede and I. Csiszar, "Common randomness in information theory and cryptography, part II: CR capacity," *IEEE Transactions on Information Theory*, vol. 44, pp. 225–240, Jan. 1998.
[12] T. M. Cover and J. A. Thomas, *Elements of Information Theory*. New York: Wiley, 1991.
[13] U. M. Maurer and S. Wolf, "Secret key agreement over a non-authenticated channel - Part I: Definitions and bounds," *IEEE Transactions on Information Theory*, vol. 49, pp. 822–831, Apr. 2003.